\documentclass{aa}  
\usepackage{graphicx}
\usepackage[varg]{txfonts}
\usepackage{natbib}
\bibpunct{(}{)}{;}{a}{}{,} 
\def\kms{km\,s$^{-1}$}

\def\msun{$M_{\odot}$}
\def\msol{$M_{\odot}$}
\def\rsol{$R_{\odot}$}

\def\s{$\sigma$}

\def\w{$\omega$}

\def\aap{A\&A}

\def\apjs{ApJS}
\def\apjl{ApJ}

\defcitealias{MGS12}{Paper~I}

\begin{document}
   \title{Three-dimensional orbits of the  triple-O stellar system HD 150136\thanks{Based on 
observations collected at the European Southern Observatory (ESO) under program IDs 089.D-0730 and 189.C-0644.}}

   \author{H.~Sana\inst{1}
          \and
          J.-B.~Le~Bouquin \inst{2} 
          \and
          L.~Mahy\inst{3}
          \and
          O.~Absil\inst{3}
          \and
          M.~De~Becker\inst{3}
          \and
          E.~Gosset\inst{3}\fnmsep\thanks{Senior Research Associate F.R.S.-FNRS}
          }

   \institute{Astronomical Institute Anton Pannekoek, 
              Amsterdam University,  
              Science Park 904, 1098~XH, 
              Amsterdam, The Netherlands\\
              \email{h.sana@uva.nl}
         \and
              UJF-Grenoble 1 / CNRS-INSU, 
              Institut de Plan{\'e}tologie et d'Astrophysique de Grenoble 
              (IPAG) UMR 5274, Grenoble, France
         \and
              D\'epartement d'Astrophysique, G\'eophysique et Oc\'eanographie,
              Universit\'e de Li\`ege, 
              All\'ee du 6 Ao\^ut 17,
              B-4000 Li\`ege, Belgium
             }

   \date{Received September 15, 1996; accepted March 16, 1997}

 
  \abstract
   {HD~150136 is a triple hierarchical system and a non-thermal radio emitter. It is formed by an O3-3.5~V + O5.5-6~V close binary and a more distant O6.5-7~V tertiary. So far, only the inner orbital properties have been reliably constrained.}
   {To quantitatively understand the non-thermal emission process, accurate knowledge of the physical and orbital properties of the object is crucial. Here, we aim to investigate the orbital properties of the wide system and to constrain the inclinations of the inner and outer binaries, and with these the absolute masses of the system components. }
   {We used the PIONIER combiner at the Very Large Telescope Interferometer to obtain the very first interferometric measurements of HD~150136. We combines the interferometric observations with new and existing high-resolution spectroscopic data to derive the orbital solution of the outer companion in the three-dimensional space. }
   {The wide system is clearly resolved by PIONIER, with a projected separation on the plane of the sky of about 9 milli-arcsec. The best-fit orbital period, eccentricity, and inclination are 8.2~yr, 0.73, and 108\degr. We constrain the  masses of the three stars of the system to $63\pm10$, $40\pm6$, and $33\pm12$~\msun\ for the O3-3.5~V, O5.5-6~V, and O6.5-7~V components. }
  {The dynamical masses agree within errors with the evolutionary masses of the components. Future interferometric and spectroscopic monitoring of HD~150136 should allow one to reduce the uncertainties to a few per cent only  and to accurately constrain the distance to the system. This makes HD~150136 an ideal system to quantitatively test evolutionary models of high-mass stars as well as the physics of non-thermal processes occurring in O-type systems. }


   \maketitle
%

\section{Introduction}

\object{HD 150136} is the brightest member of the \object{NGC 6193} cluster in the \object{Ara OB1} association and a known non-thermal radio-emitter \citep{ben2006}, i.e.\ an object where particles are accelerated to relativistic energies \citep[see][and references therein]{debrev}. This distinctive feature was the reason for a decade-long observational effort to clarify the nature and properties of HD~150136. Recently, \citet[][hereafter \citetalias{MGS12}]{MGS12} showed that HD~150136 is a triple hierarchical system consisting of an inner binary, with an O3V((f$^{\star}$))-O3.5V((f$^+$)) primary star  and an O5.5-6V((f)) secondary star  on a $P_\mathrm{in} \approx 2.67$~d orbit,  and a third physically bound O6.5-7V((f)) companion on a $P_\mathrm{out} \approx 8$- to 15-year orbit. With a total mass to be estimated around 130~\msun, HD~150136 is  one of the most massive multiple O-star systems known. It is also the closest one to Earth to harbour an O3 star.

Given the range of possible orbital periods for the outer system, \citet{MGS12} estimated a probable separation on the plane of the sky between the inner pair and the third component of roughly 10 to 40~milli-arcsec (mas). Here, we report  on extended spectroscopic monitoring that allows us to obtain the first orbital solution of the outer system. We also report on the very first interferometric detection of the wide system using the Very Large Telescope Interferometer (VLTI).


\begin{table}
\caption{Journal of the new and archival spectroscopic observations.} 
\label{tab: spectro}       
\centering              
\begin{tabular}{c r r r r} 
\hline \hline
HJD           & $v_1$ \hspace*{3mm} & $v_2$ \hspace*{3mm} & $v_3$ \hspace*{3mm} \\ 
$-$2\,450\,000 & (\kms) & (\kms) & (\kms) \\
\hline
\vspace*{-1mm}\\
\multicolumn{4}{c}{2000 UVES observation}\\
\vspace*{-1mm}\\
1726.5548 & 173.5 & $-$368.8 &  29.4      \\
\vspace*{-1mm}\\
\multicolumn{4}{c}{2008 FEROS observation}\\
\vspace*{-1mm}\\
4658.5184 & $-$107.5 & 128.2 &  27.0      \\
\vspace*{-1mm}\\
\multicolumn{4}{c}{2011 FEROS observations}\\
\vspace*{-1mm}\\
5642.916 &	$-$182.9&        268.4&	 $-$6.9\\
5696.790 &	$-$208.8&        298.5&	 $-$4.0\\
5696.907 & 	$-$191.8&        269.4&	 $-$7.7\\
5697.897 &	   186.5&     $-$337.0&	 $-$6.4\\
5699.583 &      $-$198.9&        272.9&	 $-$6.4\\
5699.588 &      $-$191.7&        268.9&	 $-$6.7\\
\vspace*{-1mm}\\
\multicolumn{4}{c}{2012 FEROS observations}\\
\vspace*{-1mm}\\
6048.661 &	   148.6&     $-$274.7&	 $-$23.7\\
6048.758 &	   120.7&     $-$221.2&	 $-$27.4\\
6048.899 &	    62.0&     $-$150.9&	 $-$29.5\\
6049.597 &	$-$215.8&	 306.1&	 $-$18.0\\
6049.715 &	$-$221.1&	 314.8&	 $-$18.0\\
6049.924 &	$-$208.2&	 280.1&	 $-$22.3\\
6050.666 &	   119.1&     $-$223.4&	 $-$26.2\\
6050.843 &	   167.7&     $-$310.1&	 $-$25.7\\
6051.641 &	    33.6&     $-$114.2&	 $-$22.1\\
6051.864 &	$-$107.5&	 128.2&	 $-$21.0\\
6052.646 &	$-$192.7&	 268.9&	 $-$20.4\\
6052.781 &	$-$152.4&	 219.6&	 $-$24.9\\
6052.909 &	$-$109.9&	 107.7&	 $-$27.6\\
6053.660 &	   188.6&     $-$344.1&	 $-$27.6\\
6053.778 &	   187.8&     $-$350.7&	 $-$28.3\\
6053.937 &	   168.4&     $-$308.4&	 $-$26.1\\
6054.649 &	$-$152.4&	 190.2&	 $-$26.5\\
6054.758 &	$-$188.9&	 240.8&	 $-$23.4\\
\hline
\end{tabular}
\end{table}

\section{Observations and data reduction}

\subsection{Spectroscopy}
We used the FEROS spectrograph mounted at the MPG/ESO 2.2m telescope at La Silla (Chile) to obtain  new high-resolution optical spectra of HD~150136 that supplement the FEROS data analysed in \citetalias{MGS12}. Eighteen FEROS spectra were obtained during a eight-night run in May 2012 (PI: Mahy), providing a continuous wavelength coverage from 3700 to 9200\AA\ at a resolving power of 48\,000. The data were processed  as described in \citetalias{MGS12}.  In addition,  we searched the ESO archives for complementary data. We retrieved six FEROS spectra from 2011 (PI: Barb\'a; 087.D-0946(A)), one FEROS spectrum from 2008 (PI: Barb\'a; 079.D-0564(B)) and one UVES spectrum from July 2000 (PI: Roueff; 065.I-0526(A)). The latter two spectra provide additional constraints on the tertiary component, but not on the systemic velocity of the inner pair and were therefore  used only to calculate the orbital solution of the wider pair. 

The  disentangling procedure described in \citetalias{MGS12} for triple systems was applied to the 2011 and 2012 sets of spectra. We also reprocessed the complete data set, using our cross-correlation technique on all data from 1999 to 2012 to consistently measure the radial velocities (RVs) of the HD~150136 components. The RV values corresponding to the new observational epochs are given in Table~1 along with the journal of the 2011 and 2012 observations.

\subsection{Long baseline interferometry}

\begin{figure*}
\centering
   \includegraphics[width=18cm]{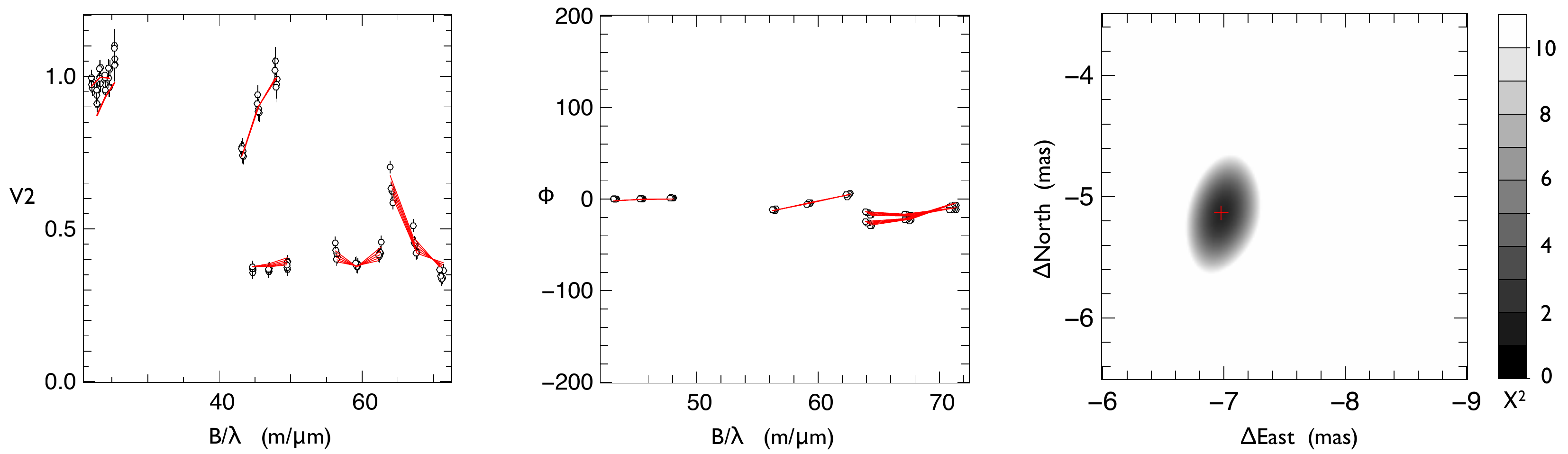}
   \caption{Calibrated visibilities (left panel) and closure phases (middle panel) from PIONIER in August 2012, overlaid with the best fit binary model (red solid lines). The right panel shows the $\chi^2$ map in the vicinity of the best-fit solution.
           }
      \label{fig:interfero}
\end{figure*}

Interferometric data were obtained in June and August 2012 with the PIONIER combiner \citep[][]{Le-Bouquin:2011} 
and the Auxiliary Telescopes (ATs) at the VLTI \citep{Haguenauer:2010}. 
Fringes were dispersed over three spectral channels across the H-band (1.50 - 1.80\,$\mu$m). 
The ATs were located in configuration A1-G1-K0-I1, providing six projected baselines 
ranging from 30 to 120~m and a maximum angular resolution of 2~mas in the H-band. Data were reduced 
and calibrated with the \texttt{pndrs} package \citep[][]{Le-Bouquin:2011}. Calibration stars were 
chosen in the JMMC Stellar Diameters Catalog \citep{Lafrasse:2010uq}.  The closure phases and 
visibilities were modelled with two unresolved sources because the expected diameters of 
the individual components ($<0.1$\ mas) as well as the separation of the inner pair ($<0.2$\ mas) 
are largely unresolved by the longest VLTI baselines. The data show no evidence that these assumptions 
may be wrong ($\chi^2_\mathrm{red}\approx1.2$). We used the 
\texttt{LITpro}\footnote{\tiny\url{www.jmmc.fr/litpro}} software \citep{Tallon-Bosc:2008} 
to extract the best-fit binary parameters, namely the flux ratio (assumed to be constant across the H-band) 
and the astrometric separation.  The closure phases were also analysed independently using the 
method presented in \citet{Absil:2011}, and the results agreed excellently.

 HD\,150136 was clearly resolved on both dates with a slight decrease in separation between June and August (Table~\ref{tab: interfero}).  Figure~\ref{fig:interfero} 
shows the data obtained in August 2012 overlaid with the best-fit binary model. 
 The final accuracy is dominated by a 2\%\ uncertainty 
in the wavelength calibration of PIONIER.  
Adopting a distance of $1.32 \pm 0.12$~kpc \citep{HeH77}, the measured angular separations translate into projected distances of  
$12.2\pm 1.1$ and $11.4\pm 1.1$~AU for the June and August observations.
The 1\s\ error-bars are dominated by the 10\%\ uncertainty on the distance.

\begin{table}
\caption{Interferometric best-fit measurements and 1\s\ error-bars.}             
\label{tab: interfero}       
\centering              
\begin{tabular}{l l l l } 
\hline\hline  & & \multicolumn{2}{c}{Observing date} \\
Parameter & Unit & 2012-06-10 & 2012-08-15 \\    
\hline 
\vspace*{-3mm}\\
HJD &($-$2\,450\,000)             & \hspace*{4mm} 6088.595 &\hspace*{4mm} 6154.573 \\
$(f_3/f_{1+2})_{1.65\mu}$ &         & \hspace*{2.5mm}$0.24 \pm 0.02$ & \hspace*{2.5mm}$0.24 \pm 0.02$   \\
$\delta$E               & (mas)   &               $-7.96 \pm 0.16$ & $-6.98 \pm 0.14$ \\
$\delta$N               & (mas)   &               $-4.73 \pm 0.09$ & $-5.13 \pm 0.10$ \\
$r$                     & (mas)   & \hspace*{2.5mm}$9.26 \pm 0.19$ & \hspace*{2.5mm}$8.66\pm0.17$ \\
$\theta$                & (\degr) & \hspace*{0.7mm}$239.2 \pm 1.0$ & \hspace*{0.7mm}$233.7\pm 1.1$ \\
$\chi^2_\mathrm{red}$    &         & \hspace*{8mm}  1.2             & \hspace*{8mm}   1.4 \\    
\vspace*{-3mm}\\
\hline                                   
\end{tabular}
\end{table}
%


\section{Orbital properties}

%
\begin{table}
\caption{Best-fit circular orbital solutions of the inner pair for the 2011 and 2012 campaigns.}             
\label{tab: RV12}       
\centering              
\begin{tabular}{l l l l } 
\hline\hline
Parameter & Unit & Primary & Secondary \\    
\hline
\vspace*{-1mm}\\
\multicolumn{4}{c}{ 2011 SB2 solution }\\ 
\vspace*{-1mm}\\
$P_\mathrm{in}$ & (d)       & \multicolumn{2}{c}{ 2.67454 (fixed) } \\
$T_0$ &(HJD$-$2\,450\,000) & \multicolumn{2}{c}{ 1327.136 $\pm$ 0.006} \\
$M_2/M_1$  &               & \multicolumn{2}{c}{ \hspace*{5.5mm}0.623 $\pm$ 0.008} \\

$\gamma$ &(\kms)       &                $-15.7 \pm 2.4$ &                $-12.1 \pm 3.2$ \\
$K$ &(\kms)            & \hspace*{.7mm}$208.3 \pm 2.3$ & \hspace*{.7mm}$334.5 \pm 3.6$ \\
r.m.s. &(\kms)         & \multicolumn{2}{c}{\hspace*{3mm} 6.2} \\
\vspace*{-1mm}\\
\multicolumn{4}{c}{ 2012 SB2 solution} \\
\vspace*{-1mm}\\
$P_\mathrm{in}$ &(d)        & \multicolumn{2}{c}{ 2.67454 (fixed) } \\
$T_0$ &(HJD$-$2\,450\,000) & \multicolumn{2}{c}{1327.151 $\pm$ 0.005 } \\
 $M_2/M_1$ &               & \multicolumn{2}{c}{ \hspace*{5.5mm}0.630 $\pm$ 0.007 } \\

$\gamma$& (\kms)       & $-20.4 \pm 2.0$ & $-12.1  \pm 2.8 $ \\
$K$ &(\kms)            & \hspace*{.7mm}$214.4 \pm 2.5$ & \hspace*{.7mm}$340.4  \pm 3.9 $ \\
r.m.s.& (\kms)         & \multicolumn{2}{c}{ \hspace*{3mm}13.1} \\

\hline                                   
\end{tabular}
\end{table}
%

\subsection{The close pair}
We used the Li{\`e}ge Orbital Solution Package\footnote{LOSP is developed and maintained by H. Sana and is available at http://www.science.uva.nl/$\sim$hsana/losp.html. The algorithm is based on the generalization of the SB1 method of \citet{wol67} to the SB2 case  along the lines described in \citet{rau00} and \citet{san06a}.} (LOSP) to compute the orbital solutions of the inner pair during the 2011 and 2012 campaigns (Table~\ref{tab: RV12}). This provided a measurement of the systemic velocity of the inner pair at these epochs. We fixed the orbital period to the same value as that of \citetalias{MGS12}, i.e.\ 2.67454~days. For both campaigns, the mass ratio, the semi-amplitudes, and the projected semi-major axis are similar to those found in Paper ~I. The difference between the systemic velocities of the primary and of the secondary components is larger for the 2012 solution but, globally, both orbital solutions of Table~\ref{tab: RV12} agree with each other and with the solution determined in \citetalias{MGS12}.


\subsection{The third companion}

\subsubsection{Spectroscopic orbital solution}
We determined the systemic velocities of the inner system  during  the individual short-term campaigns with more than one spectrum since 1999. These values are put in perspective with the RVs of the tertiary component in Fig.~\ref{fig: RV3}. There is a clear, anti-correlated periodic motion of the systemic velocity of the close binary system and of the tertiary component.  We performed a Fourier analysis on the tertiary RVs using the Heck-Manfroid-Mersch method \citep[][refined by \citealt{gos01}]{hec85}. The highest peak in the periodogram indicates a period of about $3335 \pm 260$~days. We adopted this value as our initial guess and used LOSP to refine the period estimate. We computed  an SB1 orbital solution from the RVs of the tertiary alone and an SB2 solution using the tertiary RVs and the systemic velocities of the inner pair ($\gamma_{12}$). The two solutions are in excellent agreement and the SB2 RV solution is given in Table~\ref{tab: spectro2}. The reliability of the RV-only solution, especially its eccentricity and the semi-amplitudes of the RV curves, is unfortunately limited by a lack of sampling around periastron.

\begin{table}
\caption{Best-fit LOSP RV solution of the wide system.} 
\label{tab: spectro2}       
\centering              
\begin{tabular}{l l l l} 
\hline\hline
            &            & \multicolumn{2}{c}{LOSP RV solution}   \\
Parameter    & Unit      & Inner pair       & Tertiary            \\    
\hline
\vspace*{-3mm}\\
$P_\mathrm{out}$ &(d)     & \multicolumn{2}{c}{ $2980 \pm 71$ }    \\ 
$e$             &        & \multicolumn{2}{c}{ \hspace*{3.5mm}$0.60 \pm 0.14$}   \\
\w &(\degr)              & \multicolumn{2}{c}{ $259.2\pm 5.8$}    \\
$T$ &(HJD$-$2\,450\,000) & \multicolumn{2}{c}{ \hspace*{1.7mm}$1193 \pm 104$}    \\
$M_3/M_{1+2}$ &           & \multicolumn{2}{c}{ \hspace*{3.5mm}$0.31 \pm 0.07$}   \\
$\gamma$& (\kms)         & $-21.6 \pm 2.7$ & $-16.1 \pm  4.9$   \\
$K$ &(\kms)              & \hspace*{2.2mm}$16.7 \pm 4.5$ & \hspace*{2.2mm}$53.6 \pm 14.4$   \\
r.m.s.& (\kms)           & \multicolumn{2}{c}{ \hspace*{3.mm}5.2}               \\
\hline                                   
\end{tabular}
\end{table}


\subsubsection{Simultaneous RV and astrometric orbital solution}
\begin{figure}
   \centering
      \includegraphics[width=\columnwidth]{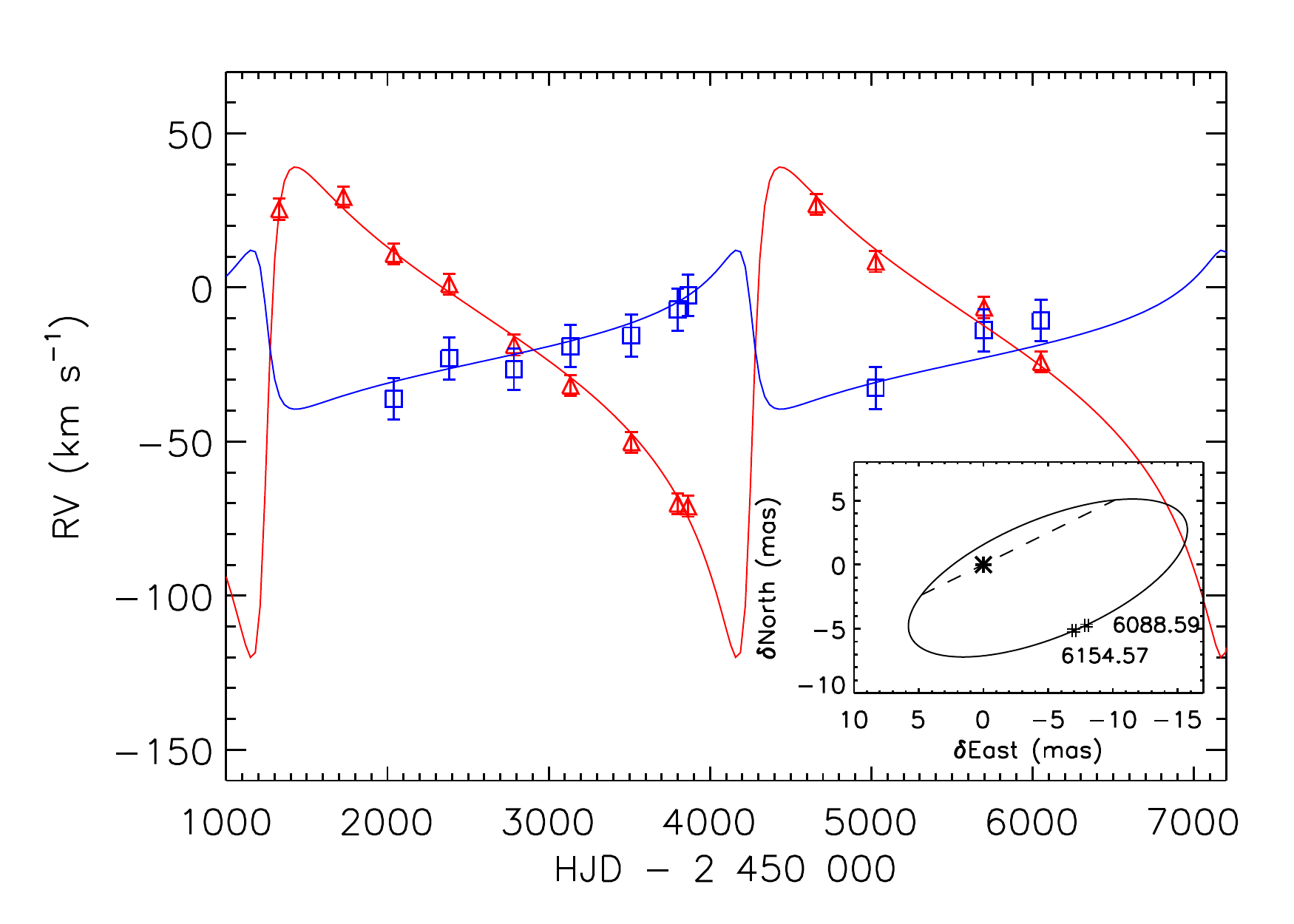}
      \caption{Main panel: Evolution of the third component RVs (triangles) and of the systemic velocity of the O+O inner system (squares) as a function of time. The best-fit RV curves are overlaid (Table~\ref{tab: final}). Insert: PIONIER astrometric points and projection of the best-fit relative orbit on the plane of the sky. The dashed line shows the line of nodes. } \label{fig: RV3}
   \end{figure}

The separations measured by PIONIER in June and August 2012 are relatively small and indicate an eccentricity at the upper end of the confidence interval of the RV-only solution of Table~\ref{tab: spectro2}. We therefore proceeded to simultaneously adjust the RV and astrometric measurements of the outer pair.  We minimized the  squared differences between the measurements and the model
\begin{eqnarray}
\chi^2& = & \sum \left( \frac{\gamma_{12} - \gamma_{12}^\mathrm{mod}}{\sigma_{12}       } \right)^2
        +   \sum \left( \frac{      v_{3} -       v_{3}^\mathrm{mod}}{\sigma_{3}        } \right)^2 \nonumber \\
      & + & \sum \left( \frac{ \delta E  -  \delta E^\mathrm{mod}}{\sigma_{\delta E} } \right)^2
        +   \sum \left( \frac{ \delta N - \delta N^\mathrm{mod}}{\sigma_{\delta N}} \right)^2
\end{eqnarray}
 using a Levenberg-Marquardt method, adopting the RV-only solution of Table~\ref{tab: spectro2} as a starting point. Based on the residuals of the tertiary  and the systemic RVs around the best-fit RV curves, we adopted $\sigma_3=3.4$ and $\sigma_{12}=6.8$~\kms\ as typical uncertainties on $v_3$ and $\gamma_{12}$.  The three-dimensional orbital solution  converges towards a higher eccentricity (although still within error-bars), hence towards larger RV curve semi-amplitudes, than the RV-only solution. All other parameters remain unchanged. The best-fit combined solution is given in Table~\ref{tab: final} and is shown in Fig.~\ref{fig: RV3}.

To estimate the uncertainties, we performed Monte-Carlo (MC) simulations. We randomly drew input RVs and astrometric positions around the best-fit orbital solution at epochs corresponding to our observing dates. We used normal distributions with  standard deviations corresponding to the $1\sigma$ measurement uncertainties. We also included the uncertainty on the distance by drawing the distance from a normal distribution centered on 1320~pc and with a standard deviation of 120~pc. We recomputed the best-fit orbital solution and inclination using 1000 simulated data sets and we constructed the distributions of the output parameters. The medians of the distributions match the best-fit values of Table~\ref{tab: final} very well. We used the distances between the median and the 0.16 and 0.84 percentiles as uncertainty estimates. The upper and lower error-bars agree within 10\%\ for all parameters; Table~\ref{tab: final} provides the average of both values.

\begin{table}
\caption{Best-fit simultaneous RV and astrometric orbital solution of the wide system. The corresponding RV curves and projected orbit on the plane of sky are displayed in Fig.~\ref{fig: RV3}. }
\label{tab: final}       
\centering              
\begin{tabular}{llll}
\hline\hline
           &             & \multicolumn{2}{c}{Combined solution}\\
Parameter  & Unit        & Inner pair       & Tertiary  \\    
\hline
\vspace*{-3mm}\\
$P_\mathrm{out}$ &(d)     & \multicolumn{2}{c}{ $3008 \pm 54$ }    \\ 
$e$             &        & \multicolumn{2}{c}{\hspace*{3.5mm}$0.73\pm0.05$ }    \\
\w &(\degr)              & \multicolumn{2}{c}{               $250.7\pm2.9$}     \\
$T$ &(HJD$-$2\,450\,000) & \multicolumn{2}{c}{                 $1241\pm38$}   \\

$M_3/M_{1+2}$   &         & \multicolumn{2}{c}{ \hspace*{3.5mm}$0.32\pm0.08$}  \\
$i_\mathrm{out}$ &(\degr) & \multicolumn{2}{c}{ $107.9\pm3.0$}     \\
$\Omega$ &(\degr)        & \multicolumn{2}{c}{ \hspace*{1.5mm}$-60\pm10$}     \\
\\
$\gamma$ &(\kms)         & $-19.9 \pm 2.2$ & $-21.5 \pm  2.0$   \\
$K$ &(\kms)              & \hspace*{2.3mm}$25.8 \pm 7.6$ & \hspace*{2.3mm}$79.8 \pm  11$   \\
$a$ &(A.U.)              & \hspace*{4mm}$5.1  \pm 1.2$ & \hspace*{2.3mm}$15.8 \pm  0.6$   \\
$M$ &(\msol)             & \hspace*{3.2mm}$102   \pm 16$  & \hspace*{5mm}$ 33 \pm  12$   \\
\\
$\chi^2_\mathrm{red}$     &    & \multicolumn{2}{c}{\hspace*{3mm} 0.94}               \\
\vspace*{-3mm}\\
\hline                                   
\end{tabular}
\end{table}

\section{Discussion}

\subsection{Physical properties}

Using the best-fit three-dimensional orbit, absolute mass estimates are $M_3=33\pm12$~\msun\ and $M_{1+2}=102\pm16$~\msun\ for the third companion and the total mass of the inner pair, respectively. Combining the total absolute mass of the inner pair with the spectroscopic minimum masses of \citetalias{MGS12} (table 4), we constrain the inclination of the inner binary to values of $49.6\pm3.6$\degr. This value is compatible with the absence of eclipse in the inner system. Best masses for the primary and secondary are thus $M_1=62.6\pm10.0$ and $M_2=39.5\pm6.3$~\msun. The agreement with the expected masses given the estimated spectral types is remarkably good \citep{MSH05}.  The dynamical masses also agree within errors with the evolutionary masses obtained in \citetalias{MGS12}: $70.6_{-9.1}^{+11.4}$, $36.2_{-1.6}^{+5.0}$ and $27.0_{-3.5}^{+3.0}$~\msun\ for components 1, 2, and 3, respectively. The measured flux-ratio in the H-band, $(f_3/f_{1+2})_{1.65\mu}= 0.24$, also agrees well with the expected H-band flux-ratio of 0.26 given the component spectral types \citep{MaP06}, which
  confirms the main-sequence nature of the tertiary object \citepalias[see discussion in][]{MGS12}.


\subsection{Non-thermal emission} \label{sect: nt}

The present study allows us to clarify to some extent the considerations on the origin of the non-thermal radio emission. As argued in Paper\,I, the synchrotron radio emission probably comes from the colliding winds in the wide orbit. The radius of the photosphere ($\tau = 1$) for the primary (whose wind dominates the free-free opacity in the system) is expected to be at most 850\,\rsol\ (at $\lambda=20$\,cm), strongly suggesting that the stagnation point of the collision zone must be located farther away (see Paper\,I). For the tertiary star, we assume two typical values for the mass loss rate: $\mathrm{\dot M}_{cl}$\,=\,10$^{-7}$\,\msol\,yr$^{-1}$ \citep[{\it classical} value,][]{muijres2012} and $\mathrm{\dot M}_{ww}$\,=\,10$^{-9}$\,\msol\,yr$^{-1}$ (more representative of the {\it weak-wind} case). We also adopt a terminal velocity of 2500\,\kms, corresponding to $2.6 \times v_\mathrm{esc}$ and $v_\mathrm{esc} \approx 970$~\kms\ for O6.5-7~V stars \citep{muijres2012}.

On these bases, we estimate that, in the classical-wind case, the stagnation point is located at about 260 and 1630\,\rsol\ from the tertiary at periastron and at apastron. In the weak-wind case, these distances reduce to about 30 and 200\,\rsol. Using the same approach as described in Paper\,I, we find that the extension of the radio photosphere of the tertiary star is shorter than the distance between the tertiary component and the stagnation point of the colliding winds, whatever the wavelength and whatever the assumption about the nature of the stellar wind (classical or weak). 

In the classical-wind case, however, the stagnation point very close to periastron could be located below the radio photosphere at longer wavelengths (13 and 20\,cm) where HD\,150136 is significantly detected, but the observations reported by \citet{ben2006} were performed in December 2003, which is far from periastron passage according to our ephemeris. In the weak-wind case,  the stagnation point is always located significantly outside the photosphere of the dense primary wind, even at periastron, and at all wavelengths. These facts are in agreement with the rather high flux densities measured between 3 and 20~cm, as reported by \citet{ben2006}.

As discussed in \citet{deb2012} for a similar system, the variability in the synchrotron radio flux from a colliding-wind binary most likely comes from two changing factors: stellar separation and free-free absorption. Following the discussion above, absorption is unlikely to dominate. Any temporal change of the  HD\,150136 flux densities would therefore be mainly attributed to the varying stellar separation along the eccentric wide orbit. One may therefore expect the radio flux density to reach its maximum very close to periastron, with a minimum at apastron. 

A radio monitoring of the wide orbit in HD\,150136 (so far non-existent) is required to validate this scenario and to achieve a more detailed description of the relevant non-thermal processes. Given the expected location of the stagnation point, the wind interaction zone may be resolvable from the stars by very large baseline radio interferometric facilities. Imaging the wind-wind collision at several epochs would help to distinguish between a weak-wind, as possibly suggested by the CMFGEN analysis in \citetalias{MGS12}, and a normal wind for the tertiary star.


\section{Summary}

We reported the very first interferometric detection of the outer companion in the hierarchical triple system HD~150136. Combining the interferometric measurements, new and archival spectroscopy with data from \citetalias{MGS12}, we obtained the first  three-dimensional orbital solution of the wider system. The best-fit solution indicates a 8.2-yr period and a high eccentricity ($e \gtrsim 0.7$). The accuracy of the PIONIER interferometric measurements allowed us to constrain the inclination of the outer orbit, and from there on,  of the inner pair to within a few degrees only. 

We constrained the masses of the three stars of the system to 63, 40, and 33~\msun\ for the O3-3.5~V, O5.5-6~V, and O6.5-7~V components. In particular, this is the first direct measurement of the mass of an early main-sequence O star. We showed that the obtained dynamical masses agree within errors with the evolutionary masses estimated in ~\citetalias{MGS12}. 

Although relative error-bars remain at the 15\%-level for masses of the inner pair components and at the 30\%-level for the tertiary mass, spectroscopic observations around the next periastron passage in 2015 and further  interferometric monitoring over a significant fraction of the orbit may provide direct mass measurements with uncertainties of a couple of per cent only. It will also allow for an accurate independent measurement of the distance to the system. This constitutes a must-do to accurately test high-mass star evolutionary models and to investigate the reality and origin of the mass discrepancy in more details \citep[e.g.\ ][]{HKV92, WeV10}. 

As for other non-thermal emitters (\object{HD 93250}, \citealt{SLBDB11}; \object{HD 167971}, \citealt{deb2012}), the present study also shows that long baseline interferometry is ideally suited to resolve the components of O-type non-thermal radio emitters. The significant advances on the three-dimensional description of the orbits in HD\,150136 is a solid base for future investigations to understand the non-thermal emission and particle acceleration processes at work in massive multiple systems.


\begin{acknowledgements}
 PIONIER is funded by the Universit\'e Joseph Fourier (UJF, Grenoble) through its Poles 
TUNES and SMING and the vice-president of research, the Institut de 
Plan\'etologie et d'Astrophysique de Grenoble, the ``Agence Nationale pour la Recherche'' 
with the program ANR EXOZODI, and the Institut National des Sciences de l'Univers (INSU) 
with the programs ``Programme National de Physique Stellaire'' and ``Programme National de Plan\'etologie''. 
The integrated optics beam combiner is the result of a collaboration between IPAG and CEA-LETI based 
on CNES R\&T funding. The authors thank all the people involved in the VLTI project. 
The use of Yassine Damerdji's orbital code is also warmly acknowledged.
This study made use of the Smithsonian/NASA Astrophysics Data System (ADS), of the Centre de 
Donn\'ees astronomiques de Strasbourg (CDS) and of the Jean-Marie Mariotti Center (JMMC). 
Some calculations and graphics were performed with the freeware \texttt{Yorick}.
\end{acknowledgements}

\bibliography{hd150136_v07}   

\end{document}